# Sesame: Self-Constructive System Energy Modeling for Battery-Powered Mobile Systems

Mian Dong and Lin Zhong
Dept. of Electrical & Computer Engineering, Rice University

**ABSTRACT**

System energy models are important for energy optimization and management in mobile systems. However, existing system energy models are built in lab with the help from a second computer. Not only are they labor-intensive; but also they will not adequately account for the great diversity in the hardware and usage of mobile systems. Moreover, existing system energy models are intended for energy estimation for time intervals of one second or longer; they do not provide the required rate for fine-grain use such as per-application energy accounting.

In this work, we study a self-modeling paradigm in which a mobile system automatically generates its energy model without any external assistance. Our solution, Sesame, leverages the possibility of self power measurement through the smart battery interface and employs a suite of novel techniques to achieve accuracy and rate much higher than that of the smart battery interface.

We report the implementation and evaluation of Sesame on a laptop and a smartphone. The experiment results show that Sesame generates system energy models of 95% accuracy at one estimation per second and 88% accuracy at one estimation per 10ms, without any external assistance. Two five-day field studies with four laptop and four smartphones users further demonstrate the effectiveness, efficiency, and noninvasiveness of Sesame.

## 1. Introduction

An energy model estimates the energy consumption by a mobile system in a given time period. Mathematically, an energy model can be represented by a function $f$ such that

$$y(t) = f(x_1(t), x_2(t), \ldots, x_n(t)),$$

where $y(t)$ is the energy consumption in time interval $t$ and $x_1(t), x_2(t), \ldots, x_n(t)$ are the system behavior data for the same period that can be read in software. To use the terminology of regression analysis, $y(t)$ is the *response* of the model and $x_1(t), x_2(t), \ldots, x_n(t)$ are the *predictors*. Most system energy models employ a linear function for $f()$. The most important metrics of a model is *accuracy* and *rate*, or the reciprocal of the time interval $t$ for which the energy consumption is modeled. While the importance of a high accuracy is obvious, a high rate is also important for many reasons. First, per-application energy accounting a multitasking system, which requires energy estimation for time intervals of 10ms or shorter or at 100Hz. Per-application energy accounting is useful not only for detecting rogue applications [1] but also for incentive mechanisms for applications that require a mobile device to consume energy for a goal beyond serving its owner, e.g., participatory sensing [2, 3] and cooperative communication [4]. Second, energy models are the foundation to energy management and optimization in operating systems (OS) [5-8] and software [9].

Existing system energy modeling approaches are however fundamentally limited in two important ways. First, all existing methods except [10-12] generate the model of a system in the lab using high quality external power measurements. Such methods are not only labor-intensive but also produce fixed energy models that are determined by the workload used in model construction. Modern mobile systems such as smartphones and laptops challenge such methods as will be analyzed in Section 3. Moreover, while cycle-accurate hardware power models exist, all reported system energy models have a low rate. That is, they estimate energy consumption for a time interval of one second or longer. In another word, they estimate energy at a rate of 1Hz or lower. While they work well for certain objectives, e.g., thermal analysis/management and battery lifetime estimation, energy models of 1Hz are fundamentally inadequate for applications such as per-application energy accounting.

Our approach to address these two limitations of existing methods is to enable a mobile system to construct a high-rate system energy model without external assistance. We call this approach *self modeling,* or *Sesame*. Instead of using special circuitry for system power measurement, e.g., [11, 13], our key idea to leverage *the smart battery interface* already available on mobile systems.

Toward realizing Sesame, we answer two technical questions in this work. First, *can battery interfaces provide the necessary accuracy and rate for the energy models from 1Hz to 100Hz?* We report an extensive characterization study of the battery interfaces of existing mobile systems, reveal their fundamental limitations in accuracy and rate, and suggest methods to overcome them. In particular, we leverage the linearity of the energy model and random nature of battery interface error to devise a model molding method that achieves high accuracy and high rate when combined with principal component analysis. Second, *how can we realize Sesame on resource-limited mobile devices?* With improved data collection support and properly scheduled model construction, we show that Sesame can be realized with negligible overhead and without intruding normal usage.

We implement Sesame for both laptop computers and smartphones. We evaluate the implementation both in lab and through two five-day field studies with four laptop users and four smartphone users, respectively, and compare the generated models against accurate external measurement. The evaluation shows that Sesame can automatically generate and adapt energy models. For a Thinkpad T61 laptop, Sesame achieves an accuracy of 88% at 100Hz and 95% at 1Hz, with a battery interface of only 0.5Hz and without any external assistance; Sesame also achieves an accuracy of 82% at 100Hz and 88% at 1Hz for a Nokia N900 smartphone with a battery interface only of 0.1Hz. The accuracy of the models generated by Sesame is comparable with that by state-of-the-art system energy modeling work employing a second system for measurement. Moreover, our two field studies of four laptop users and four smartphone users show that Sesame is able to build such energy models without interrupting users.

In summary, we make the following research contributions in building Sesame.

- A characterization of battery interfaces on modern mobile devices and their limitations.
- A suite of techniques to overcome the limitations of battery interfaces and fuel gauge ICs for self energy modeling for a rate up to 100Hz.
- The realization and evaluation of Sesame that automatically constructs accurate and fine-grain energy models using the battery interface.

The rest of the paper is organized as follows. Section 2 discusses related work in system energy modeling. Section 3 provides motivations to the proposed self-modeling approach. Section 4 reports our characterization of smart battery interfaces on commercial mobile systems. Section 5 offers the design of Sesame and its key technical components in order to address the limitations of the smart battery interfaces and reduce the overhead of self model construction. Section 6 describes the implementation details of Sesame. Section 7 evaluates the Sesame through both lab and field-based studies. Sections 8 and 9 discuss the limitations of Sesame and conclude the paper, respectively.

## 2. Related Work and Background

There is an extensive body of literature on system energy modeling. All existing system energy modeling methods give energy estimation at a low rate, such as one per minute [14], one per several seconds [6], and one per second (1Hz) [15-17]. In contrast, the goal of the proposed research is to achieve high accuracy at a rate as high as 100Hz, or one per 10ms, as required by energy accounting in multitasking systems. It is worth noting that energy models for a system *component*, e.g., processor [18, 19] and memory [20], can achieve much higher rates, often cycle-accurate. Yet such models require circuit-level power characterization, which is impractical to system energy modeling.

All existing methods except [10-12] employ external assistance, e.g., a multi-meter, to measure the power consumption of the target system. The authors of [11] leverage the special property of switching regulators to measure the energy output of a switching regulator and account the energy consumption in a wireless sensor network [13]. This approach, however, requires special circuitry that are not general available in general computer systems. Two very recent papers report solutions that build power models for virtual machines using built-in power sensors coming with servers [10] and for smartphones using the battery sensor [12]. These solutions do share our philosophy of self-modeling but falls short of achieving our goal for three important reasons. (*i*) They have a low rate, 0.5Hz [10] and 0.1Hz [12]. (*ii*) Both employ simple linear regression with *fixed* predictors, identified through careful, manual examination of the system power characteristics. In contrast, we target at adaptive models that statistically select the best predictors. Such adaptation is important to mobile devices where hardware/usage diversity is high. (*iii*) Both employ an expensive calibration to construct the model, making it inherently inefficient for adaptation.

## 3. Why Self Modeling?

We next highlight the dependencies of energy models on the configuration, usage and even the battery of a mobile system and therefore motivate the proposed self-modeling approach. Out of simplicity, we use the 1Hz energy model from [17, 21] based on CPU utilization. Our target system is a Thinkpad T61 laptop. We measure its power with a USB-2533 Data Acquisition System from Measurement Computing along with another PC.

### 3.1 Dependency on Hardware

Energy models may change due to hardware change. For example, if more memory is added, the system power consumption will likely increase, given the same CPU utilization. We study three hardware configurations of the T61 laptop with only minor differences, i.e., fixed CPU frequency of 2GHz, fixed CPU frequency of 800MHz; and DVS (dynamic voltage scaling) enabled. Similar to existing modeling methods, we build a calibrator program that switches between busy loops and idle with a controllable duty cycle to produce workload with different CPU utilizations. At each level of CPU utilization, we measure the laptop's power consumption. Figure 1 (left) presents the linear energy model for each of the hardware configura-tions. It clearly shows that there are considerable differences (up to 25%) among three models. This simple example demonstrates the generic dependency of energy models on hardware configuration. We note that the model difference in this particular example can be eliminated by adding CPU frequency as a predictor. However, it is impractical to include all predictors that can account for every hardware detail because not all hardware components are visible to the OS and collecting the values of a large number of predictors would incur significant overhead.

### 3.2 Dependency on Usage

Using the same setup, we next illustrate how usage of the system can impact its energy model. Different software may invoke different hardware components. Since some components are invisible to the OS, difference in invisible components induced by software will not be noticed by the OS. To demonstrate the usage dependency, we build three linear models by running three different benchmarks: the dummy calibrator described above, a SW Developer benchmark (Developer), and a Media Player benchmark (Player), both of the latter two from the Linux Battery Life Toolkit (BLTK) [22]. As shown in Figure 1 (right), the difference among their estimations can be as high as 25%. Again, one may argue that such difference can be reduced by incorporating as many benchmarks as possible into a factory-built model. However, using more benchmarks will lead to a model that generalizes decently but does not achieve high accuracy for the most common usage. As the usage of mobile systems by different users can be very different [23] and the usage by the same user can evolve over time [24], factory-built models are unlikely to provide an improved accuracy for all users and all usage.

### 3.3 Dependency on Battery

The energy consumption of a battery-powered system is impacted by the internal resistance of the battery, which varies from battery to battery. To illustrate this, we use two G1 smartphones (Phone 1 and Phone 2) and two batteries (Battery 1 and Battery 2) for four combinations. We build an energy model similar to that used in [16] for Phone 1 and battery 1 and test the model on all four combinations. In all the tests, we use the same benchmark described in Section 4. Table 1 provides the root mean square (RMS) of relative errors for each combination. As shown in the table, the error of the linear model is as low as 5% on Phone 1+Battery 1 where the model is built. And the model works pretty accurate, error less than 7%, on Phone 2+Battery 1. However, the error almost doubles when Battery 2 is used.

The dependencies of the system energy models on configuration, usage, and battery suggest "personalized" models be built for a mobile system, which is only possible through a self-modeling approach.

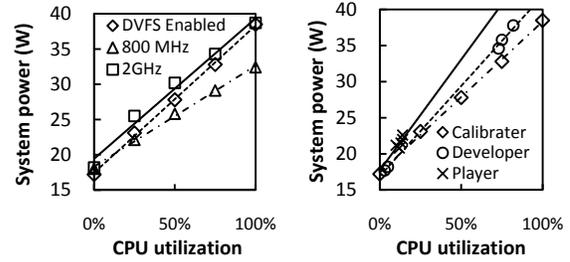

**Figure 1: Linear system energy models based on CPU utilization with (left) different configurations and (right) different usage**

**Table 1: RMS error of an energy model lab-built with Phone 1+Battery 1 for four phone-battery combinations.**

|  | Phone 1 | Phone 2 |
|---|---|---|
| **Battery 1** | 5% | 7% |
| **Battery 2** | 11% | 12% |

## 4. Battery Interface Characterization

We next provide an experimental characterization of smart battery interfaces available in modern mobile systems. We focus on their limitations in accuracy and rate in order to suggest solutions.

### 4.1 Smart Battery Interface Primer

Most batteries of modern mobile systems employ the smart battery interface in order to support intelligent charging, battery safety, and battery lifetime estimation [25].

A smart battery interface includes both hardware and software components. The key hardware of a smart battery interface is a fuel gauge IC that measures battery voltage, temperature, and maybe current. The fuel gauge IC estimates the remaining battery capacity with one of the following two methods. The first employs a built-in battery model that estimates remaining capacity using the voltage and temperature, e.g., Maxim MAX17040. The other method measures the discharge current using a sense resistor, e.g., Maxim DS2760 and TI BQ20z70. By periodically measuring the discharge current, the fuel gauge IC can count the total charges drawn from the battery and calculate the remaining capacity accordingly. Apparently, a smart battery interface with current measurement capability is much better for self energy measurement.

In software, the fuel gauge IC exposes several registers that store the values of temperature, voltage, remaining capacity, and sometimes discharge current, of the battery and updates these registers periodically. The OS, often through ACPI, provides a battery driver that can read such registers through a serial bus such as SMBus and $I^2C$. Linux employs a virtual file system for the readings from the battery driver so that applications can access battery information through a standard file system read API.

**Table 2: Battery interface of popular mobile systems**

| System | OS | Current Measurement | Reading Rate |
|---|---|---|---|
| Lenovo Thinkpad T61 | Linux | Yes | 0.5 Hz |
| Dell Latitude D630 | Linux | Yes | 1 Hz |
| Nokia N85/N96 | Symbian | Yes | 4 Hz |
| Nokia N810/N900 | Maemo | No (voltage) | ~0.1Hz |
| Google G1/Nexus One | Android | Yes | ~0.25Hz |

### 4.2 Experimental Setup

To investigate the accuracy and rate of battery interface readings, we have characterized three types of mobile systems, i.e., laptop, smartphone and PDA, as summarized by Table 2. In the characterization, we employ a set of benchmark applications for each of the mobile system. For laptops and the smartphone with Maemo, we use the BLTK [22]. For other smartphones, we use JBenchmark [26] and 3DMarkMobile [27]. Having each benchmark running, we measure the discharge current out of the battery using 1KHz sampling rate, average every ten samples and use the data, at a rate of 100Hz, as the ground truth (DAQ) to compare with the readings from battery interfaces. The measurement is performed at the output of the battery with the system being powered by the battery.

### 4.3 Key Findings

We make the following observations. *First, there is a huge variety in the smart battery interfaces on mobile systems.* We identify three types. (*i*) The voltage-based interface gives the remaining battery capacity based on the battery output voltage, e.g., Nokia N810 and N900. (*ii*) The instant interface gives the instant discharge current, such as Nokia N85 (See Figure 2 (a)). (*iii*) Finally, the filtered interface gives a low-pass filtered reading of the discharge current, such as Thinkpad T61. The fuel gauge IC used in T61 battery interface employs a low-pass filter that reports discharge current values by averaging ten readings in last 16 seconds. This low-pass filter effect is illustrated in Figure 2 (b) and is designed to facilitate the estimation of battery lifetime using history [28]. Unlike N85, the battery interface readings of T61 have a delay, around 16 seconds, than the real values. We note that modern mobile devices are increasingly using the instant interfaces. We further find that different mobile devices have different reading rates in their battery interfaces. That is, they try to give the average discharge current for time intervals of different lengths. The shorter the interval, the higher reading rate the interface has. Table 2 summarizes the battery interface reading rates of all the mobile systems we have characterized.

Consequently, our second observation is: *the rate of the smart battery interface is low with 4Hz being the highest observed.* The reason is simple: high-rate measurement requires a more expensive and more power-hungry fuel gauge IC, which does not bode well for a cost-sensitive and power-sensitive battery for mobile systems. Some fuel gauge ICs, e.g., TI BQ2019 [28], employ a charge counter, which increases by one unit whenever a certain amount of charge is drawn from the battery. As a result, the higher the current, the faster counter updates, up to 30Hz during peak power consumption. However, because a mobile system consumes moderate or low power in most of the time, the average measurement rate is only 1.138Hz according to the datasheet [28]. The low rate of smart battery interfaces is an important barrier toward our goal of realizing Sesame with high rate (up to 100Hz).

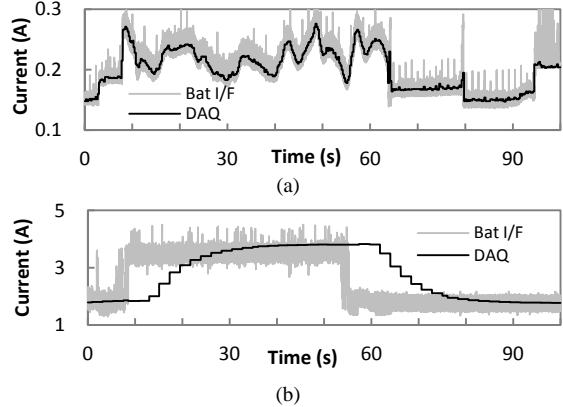

**Figure 2: Current readings from accurate external measurement and battery interface for (a) N85; (b) T61**

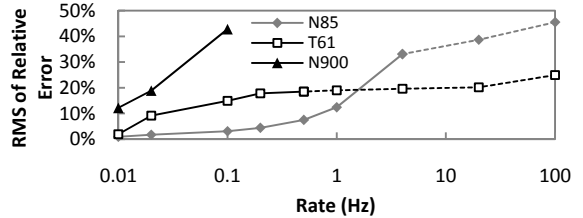

**Figure 3: Error vs. rate for battery interface readings**

Our third observation is: *the error of the instant battery interface reading is high*. We use three representative systems, N85, T61 and N900 to demonstrate this. The battery interfaces of N85 and T61 can measure current at 4Hz and 0.5Hz, respectively. That of N900 does not support direct current measurement but the average current for a period can be calculated from the readings of the remaining battery capacity at the beginning and end of the period at 0.1Hz. To evaluate the accuracy of these battery interfaces, we prepare the ground truth using the measurement from the 100Hz DAQ. For example, to compare with the 4Hz readings from N85, we use the average of the corresponding 25 samples collected from the 100Hz DAQ data as the ground truth. Figure 3 shows the root mean square (RMS) of all the relative errors for N85 at 4Hz (33%), for T61 at 0.5Hz (19%), and for N900 at 0.1Hz (45%). Such high instant errors pose a great challenge to energy modeling using smart battery interface readings.

On the other hand, we also observe that *the instant errors can be reduced by averaging readings to reduce the rate.* For example, we can average every four readings from the N85 4Hz battery interface to produce the average current for each second or a 1Hz reading. This 1Hz current reading will have only 10% RMS error compared to 33% at 4Hz. Figure 3 shows the RMS errors for N85, T61, and N900 as the reading rate is reduced down to 0.01Hz or one reading per 100s. The RMS error decreases as the reading rate decreases. There are two reasons for this error reduction. First, random noises in the smart battery interfaces contribute significantly to the instant error. By averaging the instant readings to produce a low reading, the random noise can be canceled out. Second, the low-pass filter employed by a battery interface introduces significant error, e.g., Figure 2 (b) for T61. By averaging the instant readings to reduce the reading rate, the reduced rate will approach or even drop below that of the low-pass filter's cutoff frequency, which leads to reduced error. We note that the first type of errors will be handled in model building such as linear regression thanks to their randomness; while the second type of errors are dependent on each other and should be treated as random errors. We call the second type of errors systematic errors and treat them specially, as will be described in Section 5.3.

Finally, accessing the battery interface in software contributes little to the system energy consumption in the battery interface reading. To obtain the extra energy consumed in battery interface reading, we run each of the benchmark twice. In the first time; we perform the experiment as described above, while in the second time, we only measure the power using the USB-2533 Data Acquisition System. Then we compare the average power consumption of the two and the results show that the contribution of battery interface reading is negligible, less than 1% for both systems.

# 5. Design of Sesame

We next provide the design of Sesame and our key techniques to overcome the limitations of battery interfaces

## 5.1 Design Goals and Architecture

We have the following design goals for Sesame to cater a broad range of needs in energy models.

*Sesame requires no external assistance.* It will collect the data traces, identify important predictors, and subsequently build a model out of them without another computer or measurement instrument.

*Sesame shall incur low overhead and complexity* because it runs when the system is being used. To achieve this, Sesame schedules the computation intensive tasks to run in system idle period when it is on AC power supply. Therefore, only data collection and some simple calculation are performed during system usage.

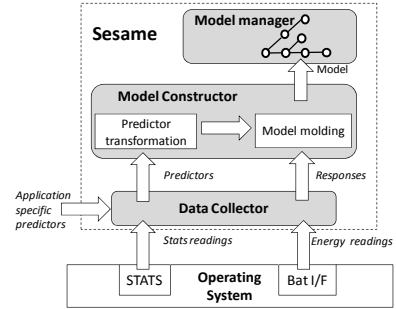

**Figure 4: Architecture of Sesame with three major components**

*Sesame should achieve higher accuracy and rate than the battery interface.* To serve a broad range of applications of energy models, Sesame should achieve high accuracy at rates as high as 100Hz. It utilizes several techniques to overcome the limitation set by the battery interface.

*Sesame must adapt to changes either in hardware or usage.* Sesame monitors the accuracy of the energy model in use and adapts it accordingly when the accuracy is below a certain threshold.

Figure 4 presents the architecture of Sesame with three major components: data collector, model constructor, and model manager, described in detail below.

## 5.2 Data Collector

The Sesame data collector interacts with the native OS to acquire predictors for energy modeling. Only the lowest layer of the data collector is platform-specific. The data collector considers the following system behavior data as widely used in existing system energy modeling work.

First, Sesame uses system statistics, such as CPU timing statistics and memory usage, supported by modern OSes. For example, Linux uses the *sys* and *proc* file system to provide information about processor, memory, disk, and process, and the *dev* file system to provide information of peripherals such as network interface card. Sesame further utilizes the Advanced Configuration and Power Interface (ACPI) available on modern mobile systems. ACPI provides platform-independent interfaces for the power state of hardware, including the processor and peripherals. In Linux, the ACPI information can be found in /proc/acpi/. Thanks to ACPI, Sesame is able to access the power status of many system components just as reading a file.

It is important to note that Sesame can incorporate any predictor that can be acquired in software. For example, an application can supply its own internal "knob" as a predictor so that Sesame can correlate the system energy consumption with the knob.

### 5.2.1 Error in Predictors

Sesame must overcome the errors in predictor readings in order to achieve a high accuracy at a high rate. There are two major sources of errors, which are more significant when reading at a higher rate.

First, some predictors are updated slowly by the OS. When the update rate is close to the reading rate, errors will occur. For example, the processor P-state residency is updated by increasing one at every kernel interrupt in Linux, 250Hz or once every 4ms by default. When predictors are collected at 100Hz or once every 10ms, there can be an error as high as 20%. Moreover, for some predictors, there exists a delay from when the predictor values are updated to when corresponding system activities are actually performed. Such predictors include the predictors related to I/O operations such as disk traffic and wireless interface traffic. For example, when the OS issues a write request to disk, the system statistics indicating sectors written to disk is immediately updated, while the write operation may be deferred by the disk driver.

Sesame employs the total-least-squares method [29], a variant of linear regression, to overcome the impact of predictor errors. While a normal linear regression minimizes the mean squares of residuals only in responses, the total-least-squares method minimizes the total squares of residuals in both responses and predictors. Numerical algorithms such as the QR-factorization and the Singular Value Decomposition (SVD) are well-known to be effective to solve such a problem [29]. Our experiments showed that the error can be reduced by about 20% by using the total-least-squares method. The time complexity of the SVD solution is $O(m^3)$.

### 5.2.2 Overhead Reduction

There is overhead for accessing the predictors. This is particularly true if Sesame is implemented out of the kernel space and each access will incur a few system calls, taking several hundred microseconds. Such overhead, along with the overhead of reading the battery interface, sets the limit of the accuracy and rate of Sesame. To reduce this overhead, Sesame employs two special design features.

First, Sesame treats predictors differently based on their update rates. For predictors that are updated faster than or comparable to the target rate, Sesame actively polls them at the target rate. These predictors include processor timing statistics and memory statistics, etc. For predictors that are updated much slower, Sesame polls them as slow as their update rates. These predictors include I/O traffic. There are also a group of predictors that are changed by user interactions such as display brightness, speaker volume, WiFi on/off switch and so on. Sesame leverages the system events produced by their updates to read them only when there is a change.

Second, Sesame implements a new system call that is able to read multiple data structures in the OS at the same time. This bundled system call also takes an input parameter that masks the predictors of no interest. This system call is one of the very few Sesame components that are platform-specific.

We note that the overhead discussed above is also true for all existing system energy modeling methods. The only additional data Sesame collects is the response, i.e., the battery interface reading. This additional overhead is very small as we measured in Section 4.3.

## 5.3 Model Constructor

The model constructor is the core of Sesame. It employs two iterative techniques, namely model molding and predictor transformation, as illustrated in Figure 4, to overcome the limitations of smart battery interfaces.

### 5.3.1 Model Molding for Accuracy and Rate

*Model molding* involves two steps to improve the accuracy and rate, respectively. The first step, called *stretching*, constructs a highly accurate energy model of a much lower rate than needed. The rationale is the third finding described in Section 4.3 that accuracy is higher for the average battery interface reading of a longer period of time and, therefore, an energy model of a low rate based on the battery interface is inherently more accurate when there exist systematic errors in battery interface readings such as the low-pass filter effect in T61. For example, Sesame will construct an energy model of 0.01Hz in the first step of model molding even if the targeted rate is 100Hz. To construct the low-rate energy model, Sesame calculates the energy consumption for a long interval by adding all the readings of the battery interface over the same interval. For example, when the rate is 0.01Hz (one reading per 100s) and the battery interface is 0.5Hz (one reading per two seconds), 50 consecutive readings of the battery interface are aggregated to derive the energy consumption for the interval of 100s. The key design issue is the choice of the low rate. A very low rate will lead to low error but it will also lead to poor generalization of the produced model, due to reduced coverage by the training data. According to our experience with many systems, we consider 0.02~0.01Hz a reasonably good range. We should note that we only need to perform model stretching when there exist systematic errors in the battery interface readings such as low-pass filter effect in T61. This is because random errors of battery interface will be handled by linear regression.

Since the first step builds a highly accurate but low-rate model, the second step of model molding *compresses* the low-rate model to construct a high-rate one. Let $\hat{y}(T) = (1, \mathbf{x}^T(T))\boldsymbol{\beta}$ denote the energy model for time intervals of $T$. $\hat{y}(T)$ is the estimated energy consumption during time interval $T$, $\mathbf{x}^T(T) = (x_1(T), x_2(T) ..., x_n(T))$ is the vector of predictors in the same interval, and $\boldsymbol{\beta} = (\beta_0, ..., \beta_n)^T$ is the vector of coefficients derived from linear regression. To estimate the energy consumption of a time interval of $t << T$, we simply collect the vector of pre-dictors for $t$, or $x^T(t)$, and use the same $\beta$ to calculate the energy consumption for the interval as $\hat{y}(t) = (1, x^T(t))\beta$.

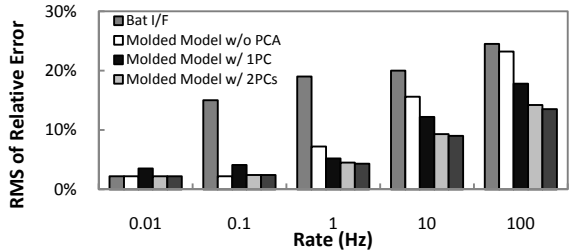

**Figure 5: Estimation errors on T61 for different time intervals and with different selections of predictors through PCA-based predictor transformation**

The rationales for this heuristic are twofold. (*i*) Ideally, the linear energy model, $\hat{y}(T) = (1, x^T(T))\beta$, holds for any length of the time interval, $T$. That is, $\beta$ should remain the same for time interval of all lengths. The Appendix provides more mathematically insight into this heuristics. (*ii*) $x^T(t)$ can be easily obtained. Many predictors are updated much faster than the response, or the battery interface reading. For example, Linux updates CPU statistics at from 250 to 1000Hz. For predictors that are not updated in $t$, their values are regarded as unchanged.

To show the effectiveness of model molding, we perform an experiment with a T61 laptop and the same setup as that used in Section 7. We build a linear model of 0.01Hz, feed it with predictors in various rates, and compare the resulted energy estimation with direct external measurements. As shown in Figure 5, at all rates from 0.01Hz to 100Hz, the accuracy of the "molded" model (Molded model w/o PCA) is always better than the battery interface (Bat/IF). The most improvement is observed for time intervals of 0.1 to 10 seconds.

Next, we will show that model molding will further improve the accuracy when combined with the second technique: predictor transformation.

### 5.3.2 Predictor Transformation

We find that principal component analysis (PCA) helps to improve the accuracy of a molded model by transforming the original predictors. PCA is not new itself but has never been explored in system energy modeling. It seeks to find a set of linear combinations of the original predictors as independent dominating factors in the predictors for the response. We explain it below

To construct the linear energy mode, we need to collect a large amount of data including both the response (energy consumption) and predictors. As defined in Section 5.2, $X(t)$ denotes the $m$ by $n$ measurement matrix of predictors in time interval $t$. Each column of $X(t)$, i.e., $x_1(t), \ldots, x_n(t)$, is a measurement vector containing $m$ measurements of predictor $x_i, i = 1, \ldots, n$. PCA transforms the $n$ predictors as follows. First, a Singular Value Decomposition (SVD) is performed on the transposed matrix of predictors $X^T(t)$, i.e. $X^T(t) = U\Sigma V^*$, where $U$ is an $n$ by $n$ unitary matrix, $\Sigma$ is $n$ by $m$ with non-negative numbers on the diagonal and zeros off the diagonal, and $V^*$ is the conjugate transpose of $V$, an $m$ by $m$ unitary matrix. The diagonal entries of $\Sigma$ are known as the *singular values* of $X(t)$ that are ranked based on their values with the highest value on the top. The rows of $U^*$, the conjugate transpose of $U$ and an $n$ by $n$ unitary matrix, are called the *singular vectors*. The higher a singular value is, the more important the corresponding singular vector is. Each singular vector defines a unique linear function that transforms the original $n$ predictors into a new predictor. Let $U_l^*$ denote the first $l$ rows of $U^*$, which are the $l$ most important singular vectors. By multiplying $U_l^*$ and $X^T(t)$, one can derive the $m$ by $l$ measurement matrix $Z(t) = (U_l^* X^T(t))^T$ for the $l$ *transformed predictors* with the $i$th column being the $m$ measurements of the $i$th new predictor. The time complexity of PCA is $O(min(mn^2, nm^2))$ or $O(m)$ since $m \gg n$.

Sesame uses the $l$ transformed predictors, instead of the original $n$ predictors, to construct the energy model as described in Section 5.3.1. By using different $l$, Sesame is able to build energy models of different accuracies. A small $l$ potentially requires fewer original predictors and, therefore, reduces the overhead of data collection, though at the cost of reduced accuracy. Sesame applies predictor transformation and model molding iteratively to reduce $l$ while maintaining a required accuracy.

Although there are still debates in the statistic community about whether PCA will help to improve linear regression in general, our results indicate that the $l$ new predictors produced by the PCA do a better job than the original $n$ predictors. Figure 5 shows that the molded model using all transformed predictors (w/all PCs) ($l = n$) is always better than the one using the original predictors (w/o PCA). The impact of $l$ on the model accuracy is also apparent. As shown in Figure 5, when $l = 2$, or only top two principal components are used, the accuracy of the model is closed to that when $l = n$ When $l = 1$, the accuracy will be highly reduced, sometimes even worse than the one without PCA. Therefore, Sesame adopts $l = 2$ in the rest of the paper unless otherwise indicated.

## 5.4 Model Manager for Adaptability

Because the model constructor may generate multiple models for different system configurations, the model manager organizes multiple energy models in a Hash table whose key is a list of system configurations. And each entry of the Hash table represents the energy model associated with a specific system configuration. Here, system configuration includes three categories: (*i*) hardware information such as component manufacturer and model; (*ii*) software settings such as DVS enabled/disabled; and (*iii*) predictors updated by user interactions such as screen brightness and speaker volume. The model manager keeps the model on the entry of current configuration as the active model. An individual model is comprised of several coefficients and the size is typically less than 1KB.

The model manager also periodically compares the energy number calculated by the model and the response value to check the accuracy. Since the battery interface is inaccurate at a high rate, the model manager checks the error of a stretched version of the active model. That is, the manager compares a very low-rate energy readings from the active energy model and from the battery interface (~0.01Hz in our implementation). The low-rate readings are obtained by aggregating the readings over ~100 seconds. If the error of the active model is higher than a given threshold, the model manager will start the model constructor to build a new one by including new predictors. Using this procedure, Sesame is able to adapt the model for usage change.

# 6. Implementation of Sesame

We have implemented Sesame for Linux-based mobile systems. The implementation works for both laptops and smartphones/PDAs using a Linux kernel. The implementation follows the architecture illustrated in Figure 4 with the model manager being the main function.

Our implementation is developed in C with around 1600 lines of codes and compiled to 320KB binary. The same code works in both platforms without any modification. The mathematical computation such as SVD is based on the GNU Scientific Library [30], which is about 800KB in binary. Sesame should work on any Linux-like system that supports GSL. Only the lowest layer of the data collector needs to be adapted for a non-Linux platform.

Sesame can be launched with user specified parameters. In particular, the user can specify rate and preferred predictors to leverage the user's prior knowledge. Sesame also provides library routines that can be used by applications of energy models in energy management and optimization. To communicate with Sesame, a program should first start Sesame while specifying the rate and the paths of the predictors. Then the program should be able to get the energy number of a time interval in the past by providing the start and end points of the interval.

### 6.1.1 Data Collector

We implement the data collector as three subroutines that are in charge of initial predictor list building, data collection and configuration monitoring, respectively. A file is used to specify the paths of predictor candidates and configuration files. By extending this library file, one can easily make Sesame work with a new platform. On T61, the data collector uses the battery interface provided by ACPI. The rate of the ACPI battery interface of T61 is 0.5Hz. On N900, the data collector talks to the HAL layer through D-Bus to access the battery interface at 0.1Hz. Table 3 provides the initial predictors used for each system.

All the data involved in the computation have only one copy and all the routines use the pointer to gain access to the data. Sesame keeps all the data traces only during predictor transformation and saves them to storage, disk or flash, when it has collected 1000 data entries. Since Sesame collects data once every 10ms during model construction, the data rate is quite low. Typically, there are about 25 predictors, all 16-bit integers, to collect for each sample, the data rate is only 5KBps and the maximal size of data trace in memory at one time is 50KB.

### 6.1.2 Model Constructor

We implement the model constructor as a subroutine that is called by the model manager. One technical challenge of implementing the model constructor is to organize data to be used for model construction in an efficient way so that response can be fast indexed given a specific set of predictors. Therefore, we choose to use hash maps to organize all the data. In a hash map, the key is a vector comprised of values of all the predictor being used, and the value is a pointer of a structure that has two members, i.e., the number and average of the responses corresponding to the predictors. When a new data sample is passed from the data collector, the model constructor first uses the predictor values to calculate an index of the hash map, finds the structure linked to the corresponding map entry, and updates the two members of the structure.

The most time consuming computation to construct a model includes PCA and linear regression. In our current implementation, the total execution time to finish the whole process takes less than one minute on T61 and about ten minutes on N900. Because neither PCA nor linear regression is needed real-time, they can easily be offloaded to the cloud. In our currently realization, they are performed when the system is wall-powered and idle.

# 7. Evaluation

In this section, we report an extensive set of evaluations of Sesame in three aspects: overhead, rate/accuracy, and the capability of adaptation using two Linux-based mobile platforms. We also report the results of two field studies to validate whether Sesame is able to generate energy models without interrupting users. For evaluation, we use the same DAQ system described in Section 4.2 to measure the system power consumption at 100Hz.

## 7.1 Systems and Benchmarks

We have evaluated Sesame using two mobile systems with representative form factors, laptop (Thinkpad T61) and smartphone (Nokia N900). Table 3 provides their specifications. As we showed in Section 4, the T61 battery interface supports current measurement at 0.5Hz; that of N900 does not support current measurement but current can be inferred for 0.1Hz. Both battery interfaces are extremely challenging to Sesame due to their low rate, especially that of N900.

**Table 3: Specifications and initial predictors of platforms used in Sesame evaluation**

|  | T61 | N900 |
|---|---|---|
| **Processor** | Intel Core 2 Duo T7100 (2GHz) | TI OMAP 3430 (600 MHz) |
| **Chipset** | Intel GM965 | N/A |
| **Memory** | DDR, 2GB | DDR 128MB |
| **Storage** | Disk 250 GB, 5400rpm | Flash 32 GB |
| **Network** | Wifi | Wifi and 3G |
| **Display** | 14.1" LCD | 3.5" LCD |
| **OS** | Linux kernel 2.6.26 | Maemo 5 |
| **Initial predictors** | CPU P-state Residency<br>CPU C-state Residency<br>Retired Instructions<br>L1/L2 Cache Misses<br>Free Memory<br>Disk Traffic<br>Wireless I/F Traffic<br>Battery Level<br>LCD Backlight Level | CPU Utilization<br>Free Memory<br>Flash Traffic<br>Wireless I/F Traffic<br>Call Duration<br>Battery Level<br>LCD Backlight Level |

**Table 4: Benchmarks used in the lab evaluation**

| BLTK | Description |
|---|---|
| Idle | System is idle while no program is running. |
| Reader | Human actions are mimicked to read local webpages in Firefox. |
| Office | Human actions are mimicked to operate Open Office. |
| Player | Mplayer plays a DVD movie |
| Developer | Human actions are mimicked to edit program files in VI and then Linux kernel is complied. |
| Gamer | Human actions are mimicked to play a part of 3D game |
| **NET** | **Description** |
| Web Browser | Human actions are mimicked to read webpages on internet in Firefox. |
| Downloader | Files of 1GB size are downloaded through SSH. |
| Video Streamer | Videos on Youtube are played in Firefox |

We choose the following three sets of benchmarks to cover diverse scenarios in mobile usage: compute-intensive (CPU2006), interactive (BLTK), and network-related (NET). CPU2006 is from SPEC. It stresses a system's processor and memory subsystem. The Linux Battery Life Toolkit (BLTK) from Less Watts [22] is designed for measuring the power performance of battery-powered laptop systems. Only three of the six BLTK benchmarks (Idle, Reader, and Developer) can run on N900. Finally, we choose three representative network applications that run on both T61 and N900 as NET. Table 4 provides the detailed information for BLTK and NET.

In experiments reported in Sections 7.2-7.4, we train and evaluate the models generated by Sesame using the same benchmarks. Therefore, the errors we will present are fitting errors as in regression terms. This evaluation is reasonable and also insightful because Sesame is intended to build models based on regular usages that are usually repetitive. Importantly, we complement this evaluation with a five-day filed study of four laptop users in which Sesame generates models using real traces in everyday use.

### 7.2 Overhead

We next show the overhead of Sesame is negligible. Figure 6 shows the overhead breakdown of self-modeling in terms of time taken for each critical stage, including response collection, predictor collection and model calculation. Since complicated computation such as PCA and linear regression is scheduled when the system is idle and being wall-powered, the model calculation only includes simple operation such as linear transformation of predictors into principal components. As shown in the figure, the total computation time of self-modeling is about 3000 μs and 5000 μs for T61 and N900, respectively, if no optimi-zation technique is used. When the data is collected at 100Hz, the overhead would account for about 30% and 50% of the period on T61 and N900, respectively, which is too high for the system.

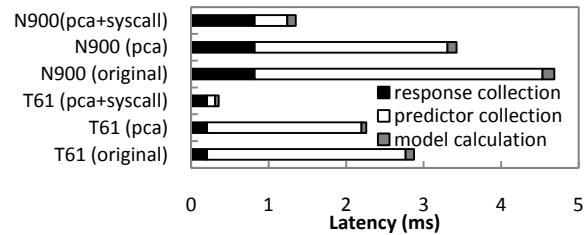

**Figure 6: Overhead of Sesame in terms of execution time. Predictor collection is most expensive among the three steps. PCA and the bundled system call (syscall) help reduce the overhead of predictor collection significantly**

We further show how the design of Sesame helps keep the overhead low. While predictor collection dominates the overall overhead, it could have been much worse without Sesame. First, predictor transformation will reduce the number of predictors such that overhead will be reduced. Second, most importantly, a bundled system call that obtains all the predictor values at one time will significantly reduce the time in predictor collection. As shown in the figure, the optimized overhead of computation time of self-modeling is about 300 μs and 1200 μs for T61 and N900, or about 3% and 12% of the period on T61 and N900, respectively. We note that the numbers above are in the worst case when response collection is performed. Actually, response collection is scheduled to operate every 2 seconds and 10 seconds on T61 and N900, respectively. Therefore, the overhead of Sesame is only 1% and 5%.

The contributions of the delay to the energy consumption can be estimated as roughly the same or slightly larger percentage-wise. Such low overhead explains the effec-tiveness of Sesame. As shown in Figure 6, the overhead of Sesame is higher for N900 than for T61. The main reason is the rate of accessing various system statistics, for both predictors and responses, is much lower on N900 than on T61. Because the system consumes extra energy for Sesame data collection, the larger overhead of N900 contributes its higher error, as will be shown later.

## 7.3 Accuracy and Rate

We next show how Sesame improves the accuracy and rate with model molding and predictor transformation.

Figure 7 shows the tradeoff curves of accuracy (in terms of the RMS of relative error) and rate (in terms of the length of the time intervals for energy estimation) for the model generated by Sesame for T610 and N900. For comparison, we also put in the tradeoff curve of the model generated using accurate external measurement with sampling rate of 100Hz, which can be also viewed as the best tradeoff curve that Sesame can possibly achieve. The relative error of each sample is calculated by comparing the estimation by an energy model with the accurate external measurement.

As shown in Figure 7, the accuracy of the Sesame on T61 is 95% at 1Hz, which is much better than the battery interface reading at the same rate. Also, Sesame is able to achieve an accuracy of 88% at 100Hz using the battery interface readings of only 0.5Hz. On N900, Sesame achieves accuracy of 86% and 82% at 1Hz and 100Hz, respectively, using the battery interface with an accuracy of only 55% at 0.1Hz. Such results highlight the effectiveness of model molding, predictor transformation, and overhead reduction by Sesame.

## 7.4 Model Convergence and Adaptation

Sesame is able to adapt the energy model to the change of configuration or usage of the computing system. We devise two experiments to test how Sesame adapts to the two types of changes with the error threshold set to be 10%. The error reported in this section is monitored by Sesame and calculated based on a rate of 0.001Hz.

In the first experiment, the T61 is running the Developer benchmark repeatedly. At the beginning, the DVS is disabled on T61 and an energy model with accuracy 92% at 1Hz, as shown in Figure 8 (Left), has been established that does not include predictors of P-state residency. Then we enable DVS, which increases the error to 17% and triggers Sesame to build a new model. In the next ten executions of the benchmarks, Sesame performs predictor transformation and includes P1 residency as an additional predictor. Finally the new energy model is able to reduce the error below the threshold (10%).

In the second experiment, T61 first runs the Developer benchmark and an energy model with accuracy 92% at 1Hz, as in experiment one. Then we switch the benchmark to the Office and the error monitored by Sesame gradually becomes worse, as in Figure 8 (Right). When the error exceeds the error threshold (10%), Sesame starts building a new energy model, which also eventually leads to an error below the threshold.

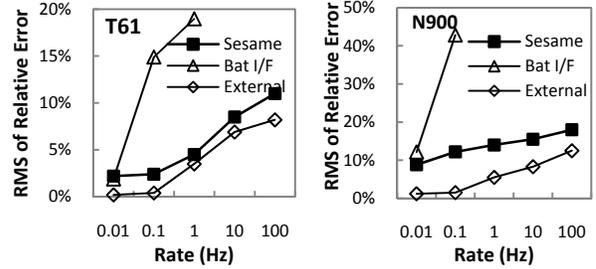

Figure 7: Model molding improves accuracy and rate. Sesame is able to increase the model rate to 100Hz

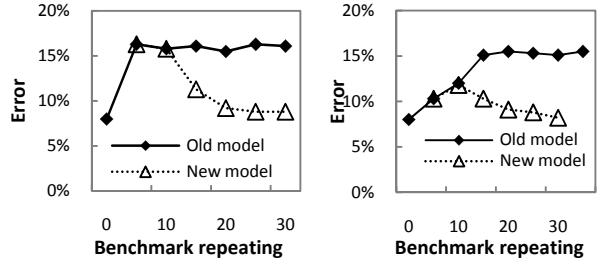

Figure 8: Model adaptation to (left) System configuration change and (right) Usage change. Error threshold is 10%

We also validate the adaptability of Sesame on N900 and observe similar results.

## 7.5 Field Study

We performed two field studies to evaluate if Sesame is able to build an energy model without interrupting normal usage and with dynamics of real field usage. The first field study involved four participants who use Linux laptops regularly. The four laptops were owned by the participants and had diversified hardware specifications, as listed in Table 5. They, however, had battery interfaces similar to that of T61 described in Section 4, i.e., with a rate of 0.5Hz and the low-pass filter effect. The second study involved another four participants who are smartphone users. Each participant was given a Nokia N900 and was asked to use the N900 as their primary phone for a week with their own SIM card in. All eight participants are ECE and CS graduate students from Rice University. The results clearly show that Sesame is able to (1) build energy models for different computing systems based on platform and usage; (2) converge in 16 hours of usage in laptops and 12 hours of usage in N900; (3) provide energy models with error of at most 12%@1Hz and 18%@100Hz for laptops and 11%@1Hz and 22%@100Hz for N900, which is comparable with state-of-the-art system energy modeling methods based on high-speed external measurement. Most importantly, Sesame achieves so without interrupting users.

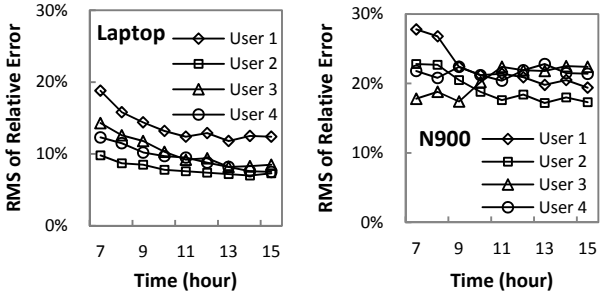
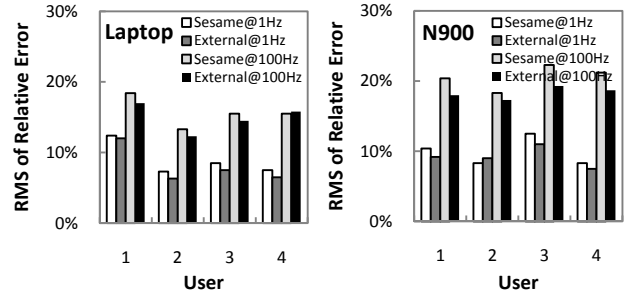

**Figure 9: Estimation errors of the models generated in the field study: model evolution in the first five days; errors are calculated relative to the averaged readings from battery interfaces in every 100 seconds in laptops and in every 1000 seconds in N900.**

**Figure 10: Comparison with readings from battery interfaces and models built base on measurements in the sixth day; errors are calculated relative to external.**

**Table 5: Specifications of laptops used in field study**

|  | User1 | User2 | User3 | User4 |
|---|---|---|---|---|
| Model | T60 | T61 | Z60m | X200s |
| CPU | Core 2 Duo T5600 @1.8GHz | Core 2 Duo T7300 @2.0GHz | Pentium M @1.6GHz | Core 2 Duo L9400 @1.8GHZ |
| MEM. | DDR2 2GB | DDR2 4GB | DDR2 1GB | DDR3 4GB |
| DISK | SATA 80GB 5400rpm | SATA 250GB 5400rpm | SATA 60GB 5400rpm | SATA 160GB 7200rpm |
| WIC | Intel 3945ABG | Intel 4965AGN | Intel 4965AGN | Intel 5300AGN |
| OS | Linux 2.6.31 | Linux 2.6.26 | Linux 2.6.31 | Linux 2.6.31 |

### 7.5.1 Procedure

Before the study, we inform the participants that the purpose of the field study is to test the properties of their batteries and they should use the mobile systems, their laptops or N900, as usual. We install Sesame on their mobile systems. To evaluate whether Sesame will affect the user experience, we set Sesame work in the first four days and stop in the last day without telling the participants. Thus, Sesame generates two energy models, with rates of 1Hz and 100Hz for the mobile system during the first four days. On the sixth day, we ask each participant to bring his/her mobile system to the lab and set up a high-speed power measurement with it. Then we ask the participants use their mobile system for 30 minutes during which Sesame is estimating energy based on the model it generated in the past five days.

We ask the participants to give a score from 1 to 5 to their experience of each day in the five-day trial regarding delay in responses. 5 means most satisfied for each of the five days in the field.

### 7.5.2 Results

Figure 9 shows the model evolvement during the modeling process. The error is calculated based on the average power readings from the battery interface in every hour. As shown in the figure, the models of all four laptop users converge in 16 hours and such convergence time for all N900 users is 12 hours. We compare the error of models generated by Sesame and that of models built from accurate external measurement. We build the linear model using predictors listed in Table 3 and the benchmarks described in Table 4. As shown in Figure 10, Sesame provides energy readings with error of at most 12%@1Hz and 18%@100Hz for laptops and 12%@1Hz and 22%@100Hz for N900, which is comparable to models built from high-speed, high-accuracy external power measurement (*External* in Figure 10).

And the results show that all the N900 users and three of the laptop users give a straight 5 for all the five days in the field study, indicating that they didn't notice any interruption in user experience. The only one participant (laptop User 2) gives two 4s for the first two days in the filed study, i.e., predictor transformation phase. The reason for the delay is he was running computation intensive applications while the initial predictor set includes ones related with performance counters which are expensive to access.

## 8. Discussions
### 8.1 Why Accuracy is Low at High Rate?

Sesame achieves a lower accuracy at 100Hz than that at 1Hz. Although the accuracy at 100Hz is still comparable to that of models built with high-speed external measurements, one may wonder: what limits the accuracy at the highest rate? We have identified three factors and discuss them below.

#### 8.1.1 Overhead of Data Collection

First, the system energy modeling is limited by the overhead of data collection for both the predictors and the response. The data collection introduces extra energy consumption and reduces the model accuracy. For an energy model of a higher rate, the predictors are collected at a higher frequency and, therefore, the energy overhead by data collection will account for a higher percentage of the measured energy. When the predictors are read at 1Hz, our experiments showed that the energy overhead only accounts for 1% of the system energy consumption. But such overhead can be as high as 20% of error when predictors are read at 100Hz. We conjecture such overhead may be calibrated and considered in the modeling to further improve the accuracy.

*8.1.2 Non-linearity*

Most existing energy modeling methods assume a linear model; Sesame indeed takes advantage of the linearity to improve the model rate. However, the relation between power consumption and system statistics is essentially non-linear. The non-linearity introduces errors when a linear model is used. In a low rate linear model, such errors can be averaged in a long time interval such that the error of the long time interval is low. We find that a non-linear model, in particular regressogram model, improves the accuracy significantly. A regressogram model is an essentially a histogram generated from the data set. Suppose there are $m$ entries in the data set and each of which includes $n$ predictors and one response. To build a regressogram model, one divides the range of each of the $n$ predictors into $k$ bins and result in $k^n$ elements. Then one puts each of the $m$ entries according to an element according to its predictor values. At last, the output of the regressogram model given a set of predictors is calculated using the average of all the entries in the same corresponding element. As shown in Figure 11, regressogram model further reduces error to only 5% in 100Hz, using the same setup as that used for Figure 7.

Unfortunately, the use of regressogram requires predictors and responses in each time interval. Even state-of-the-art battery interfaces are too slow to provide responses to build a high-rate, regressogram model.

*8.1.3 Components Invisible to Software*

Moreover, the lower-than-expected accuracy for laptop User 1 in our field study suggests another fundamental limitation of Sesame as well as all existing system energy modeling methods. Sesame and existing system energy modeling methods only employ predictors that can be collected in software or visible to the OS. Our interview with laptop User1 revealed that he used a USB disk during the study. The implementation of Sesame used in the field trial didn't differentiate the USB disk from the built-in one because the file system made the "USB" nature invisible to Sesame. Using the same trace and simulation, we found the error can be reduced to the same level as other users if the USB disk is treated separately. As recent work [31] has demonstrated, system components invisible to the OS such as the northbridge and southbridge can account for a significant portion of the overall system energy consumption. This limitation can be overcome only if future platforms

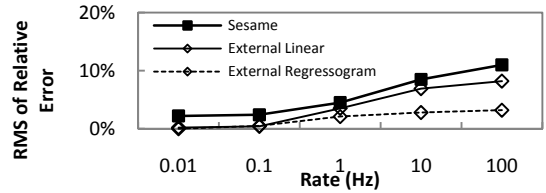

**Figure 11: Error comparison of multiple system energy models with different time resolution.**

expose more components to the OS and therefore make more predictors available to Sesame.

## 8.2 Applications of Sesame

Sesame's unprecedented ability to construct accurate, high-rate energy models *in situ* creates new opportunities in system energy optimization. We discuss a few below.

*8.2.1 System Energy Diagnosis and Management*

Sesame will enable a mobile system to automatically account the energy contribution by major hardware components and software processes, or *system energy diagnosis*. An accurate system energy diagnosis can effectively guide system energy management decisions, such as energy-driven process scheduling and intelligent power management of system components. For example, the energy diagnosis can enable accurate estimation of the critical power slope of a system [32] and inform the power manager of the best CPU rate and supply voltage to use.

Sesame will also support existing work in system energy accounting, e.g., [7, 8, 33] by providing an accurate, high-rate, and adaptive energy model. Sesame can estimate the contribution by a process or a hardware component through a linear regression analysis. Recall that Sesame builds energy models based on a set of predictors selected in the model construction process. By keeping the essential predictors for each hardware component or a group of hardware components, Sesame can construct a model in which the sum of products of the predictors and their coefficient represents the energy contribution of the corresponding hardware component.

Sesame can account energy by process using OS-reported account of each process's contribution to the value of a predictor. For example, the OS reports CPU P-state residency as contributed by an individual process. Thus, the product of each predictor and its coefficient in the linear model can be further broken down into contributions from multiple processes through the corresponding hardware component associated with the predictor.

*8.2.2 Software Energy Optimization*

Sesame can be also used to drive software energy optimization with energy models constructed through real usage by taking predictors from an individual application. Many software applications provide tunable variables or "knobs" to provide different qualities of services, either as inherent parameters for energy-aware adaptation, e.g., [34], or as variables selected by the developers for tuning the software for energy efficiency. The challenge to using such "knobs" lies in the difficulty to quantify their impact on system energy because the efficiency of an application is highly dependent on the system configuration and usage. An energy model based on such knobs are necessary [9, 35]. Sesame can readily solve this problem by incorporating such knobs as predictors in its energy construction to build an application-specific energy model only using the application knobs and therefore estimate the connections between them and the system energy directly. Because such estimations provide quantitative information regarding the energy savings by turning different knobs, the application can be adapted *in situ* either automatically by itself or by the user to effectively make tradeoffs between quality of service and efficiency.

## 9. Conclusions

In this work, we demonstrated the feasibility of self-energy modeling of mobile systems, or Sesame, with which a mobile system constructs an energy model of itself without any external assistance. We reported a Linux-based prototype of Sesame and showed it achieves 95% accuracy for laptop (T61) and 86% for smartphone (N900) with a rate of 1Hz, which is comparable to existing system energy models built in lab with the help of a second computer. Moreover, Sesame is able to achieve 88% and 82% accuracy for T61 and N900, respectively, at 100Hz, which is at least 100 times faster than existing system energy models. Our field study with four laptop users and four smartphone users further attested to Sesame's effectiveness and noninvasiveness. The high accuracy and high rate of system energy models built by Sesame allows for innovative applications of system energy models, e.g., energy-aware scheduling, rogue application detection, and incentive mechanisms for participatory sensing, on a multitasking mobile system.

Finally, although Sesame was originally intended for mobile systems, the self-modeling approach is readily applicable to other systems with self power measurement capability, e.g., servers powered by a power supply unit with a power meter. The overhead reduction and accuracy/rate improvement techniques of Sesame will be instrumental.

## 10. References

[1] H. Kim, J. Smith, and K. Shin, "Detecting energy-greedy anomalies and mobile malware variants," in *Proc. ACM/USENIX Int. Conf. Mobile Systems, Applications, and Services (MobiSys)* Breckenridge, CO, USA, 2008.

[2] S. Reddy, D. Estrin, and M. Srivastava, "Recruitment framework for participatory sensing data collections," *Pervasive Computing,* pp. 138-155, 2010.

[3] J. Burke, D. Estrin, M. Hansen, A. Parker, N. Ramanathan, S. Reddy, and M. Srivastava, "Participatory sensing," in *Workshop on World-Sensor-Web (WSW)*, 2006.

[4] K. J. R. Liu, A. K. Sadek, W. Su, and A. Kwasinski, *Cooperative Communications and Networking*: Cambridge University Press, 2009.

[5] M. Anand, E. B. Nightingale, and J. Flinn, "Ghosts in the machine: interfaces for better power management," in *Proc. ACM/USENIX Int. Conf. Mobile Systems, Applications, and Services (MobiSys)* Boston, MA, USA, 2004.

[6] F. Bellosa, A. Weissel, M. Waitz, and S. Kellner, "Event-driven energy accounting for dynamic thermal management," in *Proc. Workshop on Compilers and Operating Systems for Low Power (COLP'03)* New Orleans, Louisiana, USA, 2003.

[7] R. Neugebauer and D. McAuley, "Energy is just another resource: energy accounting and energy pricing in the Nemesis OS," in *Proc. Workshop on Hot Topics in Operating Systems (HotOS-VIII)* Oberbayern, Germany, 2001.

[8] H. Zeng, C. S. Ellis, A. R. Lebeck, and A. Vahdat, "ECO-System: managing energy as a first class operating system resource," in *Proc Int. Conf. Architectural Support for Programming Languages and Operating Systems (ASPLOS-X)* San Jose, CA, USA, 2002.

[9] J. Flinn and M. Satyanarayanan, "Energy-aware adaptation for mobile applications," in *Proc. ACM Symp. Operating Systems Principles (SOSP)* Charleston, SC, USA, 1999.

[10] A. Kansal, F. Zhao, J. Liu, N. Kothari, A. Bhattacharya, and N. Computing, "Virtual Machine Power Metering and Provisioning," in *Proc. ACM Symposium on Cloud Computing*, 2010.

[11] P. Dutta, M. Feldmeier, J. Paradiso, and D. Culler, "Energy Metering for Free: Augmenting Switching Regulators for Real-Time Monitoring," in *Proc. Int. Conf. Information Processing in Sensor Networks (IPSN)* St. Louis , MO , USA, 2008.

[12] L. Zhang, B. Tiwana, Z. Qian, Z. Wang, R. P. Dick, Z. Mao, and L. Yang, "Accurate Online Power Estimation and Automatic Battery Behavior Based Power Model Generation for Smartphones," in *Proc. Int. Conf. Hardware/Software Codesign and System Synthesis*, 2010.

[13] R. Fonseca, P. Dutta, P. Levis, and I. Stoica, "Quanto: Tracking energy in networked embedded systems," in *Proc. Symp. Operating System Design and Implementation (OSDI)*, 2008, pp. 323–338.

[14] X. Fan, W. Weber, and L. Barroso, "Power provisioning for a warehouse-sized computer," in *Proc. Int. Symp. Computer Architecture (ISCA)* San Diego, CA, USA, 2007.

[15] PowerTutor, "http://powertutor.org/."

[16] A. Shye, B. Scholbrock, and G. Memik, "Into the wild: studying real user activity patterns to guide power optimizations for mobile architectures," in *Proc. IEEE/ACM Int. Symp. Microarchitecture (MICRO-42)* New York, New York, USA, 2009.

[17] D. Economou, S. Rivoire, C. Kozyrakis, and P. Ranganathan, "Full-system power analysis and modeling for server environments," in *Proc. Workshop on Modeling Benchmarking and Simulation (MOBS'06)* Boston, MA, USA, 2006.

[18] D. Brooks, V. Tiwari, and M. Martonosi, "Wattch: a framework for architectural-level power analysis and optimizations," in *Proc. Int. Symp. Computer Architecture (ISCA)* Vancouver, British Columbia, Canada, 2000.

[19] G. Contreras and M. Martonosi, "Power prediction for intel XScale processors using performance monitoring unit events," in *Proc. ACM/IEEE Int. Symp. Low Power Electronics and Design (ISLPED)* San Diego, CA, USA, 2005.

[20] F. Rawson, "MEMPOWER: A Simple Memory Power Analysis Tool Set," IBM Austin Research Laboratory 2004.

[21] P. Ranganathan and P. Leech, "Simulating complex enterprise workloads using utilization traces," 2007.

[22] BLTK, "http://www.lesswatts.org/projects/bltk/."

[23] H. Falaki, R. Mahajan, S. Kandula, D. Lymberopoulos, R. Govindan, and D. Estrin, "Diversity in Smartphone Usage," in *Proc. ACM Int. Conf. Mobile Systems, Applications, and Services (MobiSys)* San Francisco, CA, 2010.

[24] A. Rahmati and L. Zhong, "A longitudinal study of non-voice mobile phone usage by teens from an underserved urban community," *Technical Report 0515-09, Rice University,* 2009.

[25] SBS, "http://smartbattery.org/."

[26] JBenchmark, "http://www.jbenchmark.com/."

[27] 3DMarkMobile, "http://www.futuremark.com/products/3dmarkmobile/."

[28] Texas Instruments, "SBS-Complaint Gas Gauge IC Use With The bq29311," 2002.

[29] S. V. Huffel and P. Lemmerling, *Total Least Squares and Errors-in-Variables Modeling: Analysis, Algorithms and Applications*: Dordrecht, The Netherlands: Kluwer Academic Publishers, 2002.

[30] GSL, "http://www.gnu.org/software/gsl/."

[31] R. Muralidhar, H. Seshadri, K. Paul, and S. Karumuri, "Linux-based Ultra Mobile PCs," in *Proceedings of Linux Symposium* Ottawa, Canada, 2007.

[32] A. Miyoshi, C. Lefurgy, E. V. Hensbergen, R. Rajamony, and R. Rajkumar, "Critical power slope: understanding the runtime effects of frequency scaling," in *Proceedings of the 16th international conference on Supercomputing* New York, New York, USA: ACM, 2002.

[33] F. Bellosa, "The benefits of event-driven energy accounting in power-sensitive systems," in *Proceedings of the 9th workshop on ACM SIGOPS European workshop: beyond the PC: new challenges for the operating system* Kolding, Denmark, 2000.

[34] J. Flinn and M. Satyanarayanan, "Energy-aware adaptation for mobile applications," *SIGOPS Oper. Syst. Rev.,* vol. 33, pp. 48-63, 1999.

[35] J. Lorch and A. Smith, "Software strategies for portable computer energy management," *IEEE Personal Communications,* vol. 5, pp. 60-73, June 1998.

# Appendix: Invariance of Linear Models

The key technique employed by Sesame to overcome the accuracy and rate limitations of battery interfaces is *model molding*, which first builds a linear energy model for long time intervals and then uses the same model for much shorter intervals. See Section 5.3.1. Model molding fundamentally assumes that the liner energy model is invariant at different time scales. That is, it assumes the coefficient vector $\boldsymbol{\beta}$ of the linear model, $\boldsymbol{y}(T) = (\mathbf{1}|X(T))\boldsymbol{\beta}$, is the same for $T$ of different lengths.

We next provide the physical rationale behind this heuristics. Suppose a system includes $N$ components, the $i$th component has $K_i$ states, $i = 1,2,...N$, the $i$th component in the $j$th state consumes power $P_j^i$, $i = 1,2,...N, j = 1,2,...,K_i$. Let $t_j^i$ denote the time the $i$th component spends in the $j$th state during the time interval of $T$. The energy consumption of the system during the interval can be therefore calculated as

$$y(T) = \sum_{i=1}^{N} \sum_{j=1}^{K_i} P_j^i t_j^i,$$

which is essentially a linear function of $t_j^i$. When we divide the time interval $T$ from both sides, we have

$$P = y(T)/T = \sum_{i=1}^{N} \sum_{j=1}^{K_i} P_j^i (t_j^i/T).$$

Now if we use $t_j^i/T$, the ratio of time that the $i$th component spends in the $j$th state, as predictors, the average power consumption of the system, $P = y(T)/T$, can be calculated using the same $\boldsymbol{\beta}$ no matter what time scale $T$ is in. Because it is impractical to obtain $t_j^i/T$ for all components and their states, practical energy models have to use system statistics as used in this work. Such system statistics cover major power consumers in the system and they are usually highly related to $t_j^i/T$, e.g., CPU utilization, but they usually involve the activities of multiple components and thus are correlated with each other. Therefore, the use of PCA can minimize the correlation among the system statistics and generate a new set of predictors independent with each other and closer to $t_j^i/T$. The invariance of $\boldsymbol{\beta}$ will largely hold for a linear model built with the new predictors generated by PCA.